\documentclass[aps,twocolumn,showpacs,superscriptaddress]{revtex4}

\usepackage{amsmath}
\usepackage{amsfonts}
\usepackage{amssymb}

\def\bc{\begin{center}}

\def\ec{\end{center}}
\def\be{\begin{eqnarray}}
\def\ee{\end{eqnarray}}

\newcommand{\omits}[1]{}

\def\La{\Lambda}

\def\si{\sigma}

\def\d#1#2{\frac{\displaystyle #1}{\displaystyle #2}}

\newcommand{\SR}{$SR$}

\newcommand{\SRcr}{${SR}_{c, R}$}
\newcommand{\dS}{$d{ S}$}
\newcommand{\M}{${ M}$}
\newcommand{\AdS}{${ A}d{S}$}

\newcommand{\BdS}{${ B}d{ S}$}

\newcommand{\PoR}{${ P}o{ R}$}
\newcommand{\IWR}{$ IWR$}

\newcommand{\PoIcr}{${ P}o{ I}_{c, R}$}

\newcommand{\FLT}{$FLT$}
\newcommand{\R}{$S$}
\newcommand{\E}{$E$}
\newcommand{\Lo}{$H$}

\begin{document}
\title{ Three Kinds of Special Relativity via %\\
Inverse Wick Rotation \footnote{This work is partly supported
by NSFC under %Grants
Nos.
90103004, 90403023, 10375087 and 10373003.}}

\author{ {GUO Han-Ying }}
\email{hyguo@itp.ac.cn}
\affiliation{%
   Institute of High Energy Physics, Chinese Academy of Sciences,
   P.O. Box 918-4, Beijing 100049, China,}
\affiliation{%
   Institute of Theoretical Physics, Chinese Academy of Sciences,
   P.O.Box 2735, Beijing 100080, China,}

\author{ {HUANG Chao-Guang }}
\email{huangcg@ihep.ac.cn}
\affiliation{%
   Institute of High Energy Physics, Chinese Academy of Sciences,
   P.O. Box 918-4, Beijing 100049, China,}

\author{ {XU Zhan }}
\email{zx-dmp@mail.tsinghua.edu.cn}
\affiliation{%
   Physics Department, Tsinghua University, Beijing 100084,China,}

\author{ {ZHOU Bin }} \email{zhoubihn@yahoo.com.cn}
\affiliation{%
   Department of Physics, Beijing Normal University, Beijing 100875, China.}

\begin{abstract}
Since the special relativity can be viewed as the physics in an
inverse Wick rotation of 4-d Euclid space, which is at almost equal
footing with the 4-d Riemann/Lobachevski space,
there should be important physics in the inverse Wick rotation of
4-d Riemann/Lobachevski space. Thus, there are three kinds of
special relativity in de Sitter/Minkowski/anti-de Sitter space at
almost equal footing, respectively. There is an instanton tunnelling
scenario in the Riemann-de Sitter case that may explain why $\La$ be
positive and link with the multiverse.
%\no
%Keywords: Non-Euclid geometry, inverse Wick rotation, de Sitter invariant
%special relativity
\end{abstract}


\maketitle

%%%%%%%%%%%%%%%%%%%%%%%%%%%%%%%%%%%%%%%%%%%%%%%%%%%%%%%%%%%%%%%%%%%%%%%%

In  the special relativity (\SR) in the
Minkowski-space \M$^{1,3}$ it is assumed that within a
reference frame the rigid ruler should obey the Euclid
geometry\cite{1905}. In view of field
theory, the Minkowski-space \M$^{1,3}$ is the inverse Wick rotation
(\IWR) version of 4-d Euclid space \E$^4$. Not only all geometric
issues in \E$^4$ have their physical counterparts, but there are
still additional important issues such as light-cone, simultaneity,
Einstein's formula and so on in \M$^{1,3}$ as well.

Historically, there is a well-known remarkable event in geometry on
Euclid's fifth axiom. Once it is weakened, there should be three
geometries { --- Riemann (sphere), Euclid, Lobachevski ---} at
almost equal footing \cite{Rosenfeld}. In 4-d cases, they are
defined in the space, denoted as \R$^4$/\E$^4$/\Lo$^4$, of positive,
zero and negative constant curvature of finite or infinite radius
$R$ invariant under $SO(5)$, $ISO(4)$ and $SO(1,4)$, respectively.
Remarkably, their \IWR\ versions do make sense geometrically and are
just the \dS$^{1,3}$/ \M$^{1,3}$/\AdS$^{1,3}$-space  of constant
curvature invariant under $SO(1,4)$, $ISO(1,3)$, $SO(2,3)$,
respectively.

Thus, there is a simple question:
are there such kinds of physics
in the \IWR\ version of 4-d Riemann/ Lobachevski geometry, the
\dS/\AdS-space, that should be at almost equal footing with
 \SR\ in \M$^{1,3}$?

In the present letter, we take the Riemann %sphere
geometry as an example and show that the answer is {\it yes}.

As the physics in \IWR\ version of Riemann/Euclid/ Lobachevski
geometry, there should be three kinds of \SR, denoted as \SRcr, in
\dS$^{1,3}$/\M$^{1,3}$/\AdS$^{1,3}$-space with $SO(1,4)$, $
ISO(1,3)$ and $SO(2,3)$ invariance at almost equal footing,
respectively.
In  fact, the Euclid assumption in \SR\  has less observation base in
the large scale.
Thus, weakening the Euclid assumption should lead to three kinds of
\SRcr\ based on the principle of relativity (\PoR) and the postulate
of invariant universal signal speed $c$ and length $R$ (denoted as
\PoIcr) \cite{BdS}.

In addition, there is an instanton tunneling scenario in the
Riemann-\dS\ case only that may explain why $\La$ should be positive
and link with the multiverse.

%%%%%%%%%%%%%%%%%%%%%%%%%%%%%%%%%%%%%%%%%%%%%%%%%%%%%%%%%%%%%%%%%%%%%%%%
To elucidate the above idea, we first recall the
Beltarami model as differential geometric description of the 4-d
non-Euclid geometry. We  focus on the Riemann sphere \R$^4$ and 
\dS-space.  The parallel discussion for the Lobachevski space \Lo$^4$ and 
\AdS-space is easily made.

The 4-d $S^4$ can be embedded in a 5-d Euclid space \E$^5$:%
\be\label{4s}%
{S}^4:~~\delta_{AB}\xi^A \xi^B&=&R^2, \quad A, B=0, \cdots, 4,\\\label{5ds}%
ds_E^2&=&\delta_{AB}d\xi^A d\xi^B. %
\ee%
The Beltrami model  is the intrinsic geometry of \R$^4$ in the
Beltrami space, denoted as ${\cal B}_R$, whose coordinate patch
may be set up on the one tangent to \R$^4$ at the origin $O(o^i=0)$
of Beltrami coordinates:
\be\label{Bcrd}%
x^i:=R \frac{\xi^i}{\xi^4},
\quad \xi^4\neq 0  ,\quad
 i=0,\cdots, 3,%
\ee%
with the Beltrami metric as (\ref{5ds}) restricted on ${\cal B}_R$:%
\be\label{4Bds}%
 ds_E^2&=&\{\delta_{ij}\sigma_E^{-1}(x)
 - R^{-2}\sigma_E^{-2}(x)\delta_{ik}x^k\delta_{jl}x^l\}
 dx^idx^j,\\\label{sigma}%
&&\sigma_E(x):=\sigma_E(x,x)=1+R^{-2}\delta_{ij}x^ix^j>0.
 \ee%
Obviously, to cover the whole ${\cal B}_R$, one
patch is not enough, but it is easy to check
all properties of \R$^4$ should be well-defined in the Beltrami
model patch by patch.

Since (\ref{4s}) and (\ref{5ds}) are invariant under (linear)
transformations of $SO(5)$, (\ref{4Bds}) and (\ref{sigma}) are
invariant under the fractional linear
transformations with  the same denominator, denoted by \FLT s,
among $x^i$. Their transitive form sending
a point $A(a^i)$ to the origin $O(o^i=0)$ reads,
\begin{equation}\label{FLT}
\begin{array}{l}
x^i\rightarrow
\tilde{x}^i=\pm\sigma_E(a)^{1/2}\sigma_E(a,x)^{-1}(x^j-a^j)N_j^i,\\%\nno
N_j^i=O_j^i-{ R^{-2}}%
\delta_{jk}a^k a^l
(\sigma_E(a)+\sigma_E(a)^{1/2})^{-1}O_l^i,\\
O:=(O_j^i)_{i,j=0,\cdots,3}\in SO(4).%
\end{array}\end{equation}

For two separate points $A(a^i)$ and $B(b^i)$ in ${\cal B}_R$,
\be\label{AB} %
  \Delta_{E,R}(a, b)^2
  =- R^2 \, [\sigma_E(a)^{-1}\sigma_E(b)^{-1}\sigma_E(a,b)^2-1]%
\ee %
is invariant under (\ref{FLT}).  The proper length of great `arc'
between $A$ and $B$ is
\be \label{ABL}%
L(a,b)&=& R \arcsin (|\Delta_E(a,b)|/R).%  
\ee
We may also have other issues in ${\cal B}_R$ analytically.

In view of projective geometry or simply the gnomonic projection,
the Beltrami coordinates are inhomogeneous projective ones and the
antipodal identification may be taken.   Then the great circles in
(\ref{4s}) are mapped to  straight lines, exactly
the geodesics in ${\cal B}_R$,  and vice versa.

%%%%%%%%%%%%%%%%%%%%%%%%%%%%%%%%%%%%%%%%%%%%%%%%%%%%%%%%%%%%%%%%%%%%%%%%
Now, we take an \IWR. It changes the metrics from $(\delta_{AB})$,
$(\delta_{ij})$ %
to $ (\eta_{AB}):=diag(-1,1,{ 1, 1},1)$, $(\eta_{ij}):=diag(-1,1,1,1)$. %
Then both $\xi^0$ and $x^0$ become time-like. In order to
introduce the time coordinate, a universal constant $c$ of speed
should be introduced, say $x^0=ct$. Thus, there are two
universal constants $c$ and $R$.

Via such an \IWR, the \R$^4 \subset$ \E$^5$ and the Beltrami-space
${\cal B}_R$ turn to the \dS-hyperboloid ${\cal H}_R$
in 5-d Minkowski-space \M$^{1,4}$ and the \dS-space with Beltrami metric
(\BdS-space), respectively. But, in order to preserve the
orientation, the antipodal identification should not be taken.

The \dS-hyperboloid ${\cal H}_R \subset$ \M$^{1,4}$ reads:
 \be\label{5sphr}%
 {\cal H}_R:  &&\eta^{}_{AB} \xi^A \xi^B= R^2,%
\\ %
\label{ds2}%
&&ds^2=\eta^{}_{AB} d\xi^A d\xi^B , %~
\ee
$R$
may be identified with the cosmological constant, $R^2:=3\Lambda^{-1}$.
Clearly,  Eqs. (\ref{5sphr}) and (\ref{ds2}) are invariant under
\dS\ group $SO(1,4)$. Then the Beltrami metric in \BdS-space reads
\begin{eqnarray}\label{bhl}%
 ds^2&=&[\eta_{ij}\sigma(x)^{-1}- R^{-2}
\eta_{ik}\eta_{jl}x^k x^l \sigma(x)^{-2}]dx^i dx^j,\\
\label{domain} &&\sigma(x)=\sigma(x,x):=1+R^{-2} \eta_{ij}x^i
x^j>0.
\end{eqnarray}
$\sigma(x)=0$ is the (projective) boundary of  \BdS,  denoted by
$\partial$\BdS. Under the \IWR-version of (\ref{FLT}),
\begin{equation}\label{G}
\begin{array}{l}
x^i\rightarrow \tilde{x}^i=\pm\sigma(a)^{1/2}\sigma(a,x)^{-1}(x^j-a^j)D_j^i,\\
D_j^i=L_j^i -{ R^{-2}}%
\eta_{jk}a^k a^l (\sigma(a)+\sigma(a)^{1/2})^{-1}L_l^i,\\
L:=(L_j^i)_{i,j=0,\cdots,3}\in SO(1,3),
\end{array}\end{equation}
(\ref {bhl}) and (\ref{domain}) are invariant.

The \IWR-version of (\ref{AB}) in \BdS,
\be\label{lcone0} %
 \Delta_R(a, b)^2 = -R^2\,[\sigma(a)^{-1}\sigma(b)^{-1}\sigma(a,x)^2-1]%
\ee %
is again invariant under the \FLT s of $SO(1,4)$. Thus, the
`great arcs' between $A$ and
$B$ are classified as timelike, null or spacelike, according to%
\begin{equation}\label{lcone}%
\Delta_R^2(a, b)\lesseqgtr 0.%
\end{equation}
The proper lengths of time/space-like `great arcs'
 between $A$ and $B$ are then
\be \label{AB1}%
S_{t-like}(a, b)&=&R \sinh^{-1} (|\Delta(a,b)|/R), \\
\label{AB1sl} S_{s-like}(a,b)&=& R \arcsin (|\Delta(a,b)|/R).%
\ee

Again, via the `gnomonic projection'
the timelik/null/spacelike `great arcs' are mapped to timelike/null/spacelike
straight lines in \BdS-space, respectively.   In particular, the
timelike `great arcs'  describe the uniform motions for the law of
inertia.  Then the Beltrami-coordinates turn into the inertial
coordinate systems in \BdS-space.  The property 
is just for a headstone of the \PoR.

Thus, the \IWR-counterpart of the Beltrami model of a 4-d Riemann
space \R$^4$ strongly indicates that the \dS-invariant \SR\ should
be set up in \BdS-space.

%%%%%%%%%%%%%%%%%%%%%%%%%%%%%%%%%%%%%%%%%%%%%%%%%%%%%%%%%%%%%%%%%%%%%%%%
To set up \dS-invariant \SR, the \PoR\ should be first re-stated in
the following way.  There exist a set of
inertial coordinate systems, in which the free particles and
light-signals move uniformly along straight lines, the laws of
nature without gravity are invariant under the transformations
among them.  To replace the principle of invariant light speed,
We propose the \PoIcr : In the inertial
reference frames, there exist two invariant universal constants
--- speed $c$ and length $R$, which are the local speed of light
in vacuum at the origin and curvature radius, respectively.
Based upon the \PoR\ and the \PoIcr, the \dS-invariant \SR\ can
be set up in \BdS-space. The detailed discussion can be found in
\cite{BdS}.  Here, only some main significance is  summarized.

There is a {\it law of inertia} in \BdS. Both the free
particles with rest mass $m$ and light signals without undergoing any unbalanced forces should keep their uniform motions. The motion is
known as the inertial motion.

For an inertial motion of particles, the
weighted 4-momentum $p^i=m\si^{-1}dx^i/ds$ and 4-angular-momentum $L^{ij}=x^ip^j-x^jp^i$, a weighted anti-symmetric tensor, are conserved.  They satisfy the
generalized Einstein's famous formula:%
\be \label{eml}
-\eta_{ij}p^i p^j - \frac{1}{2R^2}\eta_{ik}\eta_{jl}L^{ij}L^{kl}=m^2
\ee

For the forced motions, the second law of mechanics can be written as
\be \frac{dp^i}{ds} = f^i, \qquad \frac{dL^{ij}}{ds}=M^{ij}, \ee
where $f^i$ is the weighted 4-force and $M^{ij}=x^if^j-x^jf^i$ is the
4-moment.  If one incorporates $p^i$ and $L^{ij}$ into 5-angular
momentum ${\cal L}^{AB}$
and $f^i$ and $M^{ij}$ into 5-moment ${\cal M}^{AB}$
such that
\be
&p^i=R^{-1}{\cal L}^{4i}=-R^{-1}{\cal L}^{i4},  \quad
&L^{ij}={\cal L}^{ij}, \\
&f^i=R^{-1}{\cal M}^{4i}=-R^{-1}{\cal M}^{i4},  \quad
&M^{ij}={\cal M}^{ij},
\ee
then the second law will take the form
\be
\d {d{\cal L}^{AB}} {ds} ={\cal M}^{AB}.
\ee

There are two kinds of simultaneity in \dS-invariant \SR,
Beltrami time coordinate simultaneity and proper time simultaneity.
The Beltrami time coordinate simultaneity is responsible for the
inertial motion, inertial law, inertial reference frame, inertial
coordinate system, etc.  The proper time
simultaneity is closely linked with the cosmological
principle and comoving observers. In fact,
if  the proper time of the inertial observer at the spatial origin is taken as a `cosmic time', the Beltrami metric
(\ref{bhl}) becomes a Robertson-Walker(RW)-like metric with a
positive spatial curvature.  It shows that the 3-d cosmic space is $S^3$ rather than
flat. The deviation from the flatness is of order $\Lambda$. Our
universe might be asymptotically expanding to it.

The two kinds of simultaneity do make sense in different types of
observers' measurements. The first concerns the inertial observers
${\cal O}_I$'s measurements in their laboratories. The second
concerns with cosmological observations of co-moving observers
${\cal O}_C$.  In fact, the set of observers $\cal O$ become the
inertial-co-moving observers ${\cal O}_{I-C}$ with two-time-scale
timers. Referring to  the Beltrami time but not the proper time,
they are inertial, while inversely they are co-moving. Further, it
is very meaningful that the relation between the Beltrami metric
with coordinate time $x^0$ and its RW-like counterpart with cosmic
time $\tau$. It links the \PoR\ in \BdS-space with the cosmological
principle in its RW-like version.

The above discussions are easily generalized to the Lobachevski
space \Lo$^4$ and \AdS-invariant-\SR.  In fact, the original
Beltrami model \cite{beltrami} is just for the  Lobachevski plane.

%%%%%%%%%%%%%%%%%%%%%%%%%%%%%%%%%%%%%%%%%%%%%%%%%%%%%%%%%%%%%%%%%%%%%%%%

In view of general relativity, the curvature of
\BdS-space is gravity. In view of \dS-invariant \SR, however, there is no
gravity in \BdS-space.

Thus, how to describe the gravity is to be studied further. In
view of the gauge principle and the principle of equivalence in
the sense of \dS-invariance, the gravity might be local
\dS-invariant with local \BdS-space anywhere and anytime in the
universe.

In this sense of local \dS-gravity \cite{dSg} and its conformal
extension \cite{CG}, the Beltrami-space ${\cal B}_R$ may be regarded
as a gravitational instanton  of Euler number $e=2$ with \BdS-space
as its \IWR-version.  Since both the 4-d Euclid and Lobachevski
spaces \E$^4$ and \Lo$^4$ have no 4-d nonzero topological numbers,
this may provide a tunneling scenario { from nothing}
for the Riemann-\dS\ case only with an {\it ensemble} of \BdS-spaces of
different cosmological constants.  Although for the single instanton it is
similar to the picture of quantum cosmology \cite{Hawking},
here it may explain why $\La$ should be positive and also link with
the multiverse (see, e.g.\cite{tegmark}).  We will explain it in detail elsewhere.

%%%%%%%%%%%%%%%%%%%%%%%%%%%%%%%%%%%%%%%%%%%%%%%%%%%%%%%%%%%%%%%%%%%%%%%%

In conclusion, there should be three kinds of \SR\ as the
physics of the \IWR\ version of the Riemann/Euclid/ Lobachevski
geometry at almost equal footing, respectively. Both the
\dS/\AdS-invariant \SRcr\ can be set up based on the \PoR\ and
\PoIcr. Their most properties are in analogous with
Einstein-Poincar\'e's ${SR}$ except there is a proper room of the
curvature radius $R$ or the cosmological constant $\La$ and coincide
with ${ SR}$ if $\La\to 0$.  In fact, all local experiments
allow there exist three kinds of \SR\ at almost equal footing.
Recent observations on the Univserse, however, imply that the
\dS-invariant \SR\ with tiny positive cosmological constant may pay
an important role  as a foundation of physics in the large scale.

In addition, for the Riemann-\BdS\ case, there is an instanton tunneling
scenario from nothing. It may explain why $\La$ be positive and link with the
multiverse.\\

%%%%%%%%%%%%%%%%%%%%%%%%%%%%%%%%%%%%%%%%%%%%%%%%%%%%%%%%%%%%%%%%%%%%55

We would like to thank Professor Q.K. Lu for valuable discussions.

\end{document}